\documentclass[prd,nofootinbib]{revtex4}
\usepackage{graphicx}
\usepackage{amsmath} 
\usepackage{amssymb}
\usepackage{bm}
\def\be{\begin{equation}}
\def\ee{\end{equation}}
\def\ba{\begin{eqnarray}}
\def\ea{\end{eqnarray}}

  \def\be{\begin{equation}}
\def\ee{\end{equation}}
 \def\bi{\begin{itemize}}
 \def\ei{\end{itemize}}
  \def\ben{\begin{enumerate}}
\def\een{\end{enumerate}}
  \def\bt{\begin{tabular}}
\def\et{\end{tabular}}
\def\bc{\begin{center}}
\def\ec{\end{center}}
\def\la{\label}
\def\kap{\kappa}
\def\gam{\gamma}
\def\bea{\begin{eqnarray}}
\def\eea{\end{eqnarray}}
\def\l{\left}
\def\r{\right}
\def\f{\frac}
\def\hub{{\cal H}}
\def\d{\partial}
\def\dRoY{\left( \frac{\delta R}{Y} \right)}   
\newcommand{\koa}[1]{\frac{k^{#1}}{a^{#1}}}			

\def\le{\left}
\def\ri{\right}
\def\fr{\frac}
\def\la{\label}
\def\kap{\kappa}
\def\gam{\gamma}

\def\frru{\frac{f_{RR}}{(1+f_{R})}}

\begin{document}

\title{Dynamics of Linear Perturbations in $f(R)$ Gravity}
\author{Rachel Bean$^{1}$}
\author{David Bernat$^{1}$}
\author{Levon Pogosian$^{2,3}$}
\author{Alessandra Silvestri$^{3}$}
\author{Mark Trodden$^{3}$}

\affiliation{$^{1}$ Department of Astronomy, Cornell University, Ithaca, NY 14853, USA}
\affiliation{$^{2}$Department of Physics, Simon Fraser University, Burnaby, BC, V5A 1S6, Canada}
\affiliation{$^{3}$Department of Physics, Syracuse University, Syracuse, NY 13244, USA}
\date{\today}

\begin{abstract}
We consider predictions for structure formation from modifications to general relativity in which the Einstein-Hilbert action is replaced by a general function of the Ricci scalar. 
 We work without fixing a gauge, as well as in explicit popular coordinate choices, appropriate for the modification of existing cosmological code. We present the framework in a comprehensive and practical form that can be directly compared to standard  perturbation analyses.

By considering the full evolution equations, we resolve perceived instabilities previously suggested, and instead find a suppression of perturbations. This result presents significant challenges for agreement with current cosmological structure formation observations.  

The findings apply to a broad range of forms of $f(R)$ for which the modification becomes important at low curvatures, disfavoring them in comparison with 
the $\Lambda$CDM scenario. As such, these results provide a powerful method to rule out a wide class of modified gravity models aimed at providing an alternative explanation to the dark energy problem. 
\end{abstract}

\maketitle
\section{Introduction}
Approaches to resolving the late-time acceleration of the universe may be divided into three broad 
 classes  (see, for example,~\cite{Bean:2005ru} for a brief review). Perhaps the simplest possibility is that there is some, as yet undiscovered, consequence of our existing model of gravity and matter that leads to acceleration during the current epoch. Included in this category is the existence of a tiny cosmological constant and the possibility that the backreaction of cosmological perturbations might cause self-acceleration. A second option is the idea that a new dynamical component exists in cosmic energy budget. Included here are sources of energy density modeled by a scalar field, usually referred to as {\it dark energy}. Of equal interest, however, is that General Relativity (GR) can be modified in the low curvature regime, to admit self-accelerating solutions in the presence of negligible matter~\cite{Dvali:2000hr,Deffayet:2000uy,Deffayet:2001pu,Freese:2002sq,Dvali:2003rk,Carroll:2003wy,Capozziello:2003tk,Vollick:2003aw,Flanagan:2003rb,Flanagan:2003iw,Vollick:2003ic,Soussa:2003re,Nojiri:2003ni,Carroll:2004de,Arkani-Hamed:2003uy,Gabadadze:2003ck,Moffat:2004nw,Clifton:2004st,delaCruz-Dombriz:2006fj,Carroll:2006jn}.

We consider, in this paper, the class of modified gravity models in which the gravitational action contains a general function $f(R)$ of the Ricci scalar. For such models, the analysis of the background cosmological evolution can be significantly simplified by performing a conformal transformation on the metric.  Such a transformation maps from a frame in which the gravitational action and resulting field equations are modified from GR, commonly called the Jordan frame, to a frame in which the gravitational action for the new metric is the Einstein-Hilbert one, commonly called the Einstein frame. In this new frame, the matter fields couple non-minimally to the new metric (matter no longer falls along geodesics of the new metric) and an extra degree of freedom now manifests as a new scalar field. These couplings affect the evolution of perturbations in a potentially observable way~\cite{Bean:2001ys}.

This paper focuses on structure formation in general $f(R)$ theories and comparison of the predictions with observations. This complements previous work investigating the background evolution \cite{Carloni:2004kp,Amendola:2006mr,Capozziello:2006dj,Brookfield:2006mq} and solar system implications \cite{Brookfield:2006mq,Rajaraman:2003st,Olmo:2005jd,Allemandi:2005tg,Multamaki:2006zb,Faraoni:2006hx,Ruggiero:2006qv,Erickcek:2006vf} for $f(R)$ modifications. As part of this analysis we derive the linear cosmological perturbation theory for a general form of $f(R)$ considering only scalar perturbations, since the tensor and vector modes are unaffected by $f(R)$ modifications to gravity. We present our equations without fixing a gauge, as well as in two explicit coordinate choices: the conformal Newtonian and synchronous gauges. These equations are presented in a comprehensive and practical form that can be directly compared to standard perturbation analyses (e.g.~\cite{Ma:1994dv, Kodama}) and are applicable to the modification of existing cosmological codes such as CAMB~\cite{camb}.  This work complements alternative formulations in the Palatini approach~\cite{Koivisto}, for the metric approach in the Einstein frame~\cite{Hwang-Noh} and, more recently, in terms of  ``frame-independent" variables~\cite{Catena:2006bd}. 

In section~\ref{fofrdescription} we give an overview of the $f(R)$ action and in section~\ref{conformal} discuss the conformal transformation used to express $f(R)$ theories in a frame with Einstein field equations with non-minimally coupled matter.  In sections \ref{perttheory} and \ref{evol} we present the main analytical results of the paper. In section ~\ref{perttheory} we present full perturbation equations for $f(R)$ theories in the Jordan frame. In section~\ref{evol} we examine the general behavior of late time structure formation in both Einstein and Jordan frames specifically for $f(R)$ theories that allow late-time cosmological acceleration solutions. This leads to important constraints on a large class of $f(R)$ theories dominating in the low curvature regime. For concreteness we present the predictions for two specific examples. We also outline the origin of the apparent zeroing of density perturbations when approximating the evolution equations in the Jordan frame \cite{Zhang:2005vt}, and resolve the matter by considering the full Einstein frame equations. Finally in section~\ref{conclusions} we summarize our findings and discuss implications.

\section{Description of $f(R)$ Gravity Models}
\label{fofrdescription}

The class of theories on which we focus has action in the Jordan frame 
\be
S=\frac{1}{2\kappa^{2}}\int d^4 x\sqrt{-g}\, \left[R+f(R)\right] + \int d^4 x\sqrt{-g}\, {\cal L}_{\rm m}[\chi_i,g_{\mu\nu}] \ ,
\label{jordanaction}
\ee
where $\kappa^{2}= 8\pi G$ and the function $f(R)$ is a general function of the Ricci scalar, $R$. The matter Lagrangian, ${\cal L}_{\rm m}$, is minimally coupled and therefore  the matter fields, $\chi_i$,  fall along geodesics of the metric $g_{\mu\nu}$. The field equations obtained from varying the non-minimally coupled gravity action~(\ref{jordanaction}) with respect to $g_{\mu\nu}$ are
\be \label{jordaneom}
\left(1+f_R \right)R_{\mu\nu} - \frac{1}{2} g_{\mu\nu} \left(R+f\right) + \left(g_{\mu\nu}\Box 
-\nabla_\mu\nabla_\nu\right) f_R =\kappa^{2}T_{\mu\nu} \ ,
\ee
where we have defined $f_R\equiv \partial f/\partial R$. We assume matter to behave as a perfect fluid, with energy-momentum tensor
\begin{equation}
T_{\mu\nu} = (\rho+ P)U_{\mu} U_{\nu} + p g_{\mu\nu}\ ,
\label{perfectfluid}
\end{equation} 
where $U^{\mu}$ is the fluid rest-frame four-velocity, $\rho$ is the energy density and $P$ is the pressure. We relate $P$ and $\rho$ via $p=w\rho$, where $w$ is the equation of state parameter (for pressureless matter $w=0$ and radiation to $w=1/3$).

When considering the background cosmological evolution, we take the metric to be of the
flat Robertson-Walker form, $ds^2 = a^{2}(\tau)(-d\tau^{2} + d{\bf x}^2)$, with $a(\tau)$ the scale factor and $\tau$ conformal time, in terms of which the curvature scalar satisfies $R =6 a''/a^{3}$. Here a prime denotes differentiation with respect to conformal time, $\tau$, $\hub \equiv a' / a$ is the equivalent Hubble expansion rate.  

In the Jordan frame, the metric is that of standard GR minimally coupled to matter. Hence the stress tensor and its conservation laws will remain the ones of standard GR. 
The continuity equation is the usual 
\ba
\rho'+3\hub(\rho+P) &=& 0 \ .
\label{jordancontinuity}
\ea
However, since we have modified the gravitational action, there are extra terms appearing in the Einstein equations. In particular, for our cosmological ansatz, the Friedmann equation becomes
\ba
(1+f_{R})\hub^2+\frac{a^{2}}{6}f-\f{a''}{a}f_{R} +\hub f'_{R} &=&  \f{\kappa^{2}}{3}a^{2}\rho
\label{jordanfriedmann}
\ea
and the acceleration equation is
\bea
\f{{a}''}{a}-(1+f_{R})\hub^{2}+a^2\frac{f}{6}+\hub f'_{R}+\f{1}{2}f_{R}''&=& -\frac{\kappa^{2}}{6}a^{2}(\rho+3P) \ .
\eea

\section{Mapping to the Einstein Frame}
\label{conformal}

Although the background cosmological behavior can be chosen by an appropriate choice of $f(R)$, this freedom comes at the expense of unfamiliar dynamical equations.  There exists, however, a complementary, and sometimes conceptually simpler, way in which to approach $f(R)$ modifications to GR.  It is possible to perform a conformal transformation on the metric so as to render the gravitational action in the usual Einstein Hilbert form of GR.  The price one pays for this simplification is a non-minimal coupling between matter fields and the new metric \cite{Cotsakis:1988,Amendola:1999er,Bean:2000zm}, as well as the appearance of a new scalar degree of freedom evolving under a potential determined precisely by the original form of the $f(R)$ coupling in the Jordan action. 

Using the approaches of Chiba~\cite{Chiba:2003ir} and of Magnano \& Sokolowski~\cite{Magnano:1993bd}, following \cite{Cotsakis:1988}, we recast the gravitational action~(\ref{jordanaction}) into a dynamically equivalent form by introducing an intermediate scalar field $\Phi$.  The equivalent action is \cite{Magnano:1993bd}
\ba
S&=&\frac{1}{2\kappa^{2}}\int d^4 x\sqrt{-g}\, \left[ (\Phi + f(\Phi)) + (1 + f_{\Phi})(R - \Phi) \right]   \ \ \ \ \nonumber
\\ &&+\int d^4 x\sqrt{-g}\, {\cal L}_{\rm m}[\chi_i,g_{\mu\nu}] \ ,
\label{IntermediateAction}
\ea
where $f_{\Phi} \equiv \partial{f}/\partial{\Phi}$.  One can verify that, if $d^{2}f/d\Phi^{2} \neq 0$, the field equation for $\Phi$ is $R=\Phi$, which reduces~(\ref{IntermediateAction}) to the original action. Next consider the conformal transformation
\be
{\tilde g}_{\mu\nu} = e^{2 \omega(x^{\alpha})} g_{\mu\nu} \ ,
\label{conftrans}
\ee
such that the function $\omega(x^{\alpha})$ satisfies
\be
e^{-2 \omega}(1+f_{R}) = 1 \ .
\label{constraint}
\ee

With this choice of $\omega$ the action~(\ref{IntermediateAction}) transforms into an action with the usual Hilbert-Einstein form for gravity. If we now define the scalar field  $\phi \equiv 2\omega / \beta\kappa$, where $\beta \equiv\sqrt{2/3}$, the resulting action becomes 
\ba
{\tilde S}&=&\frac{1}{2\kappa^{2}}\int d^4 x\sqrt{-{\tilde g}}\, {\tilde R} 
+\int d^4 x\sqrt{-{\tilde g}}\, 
\left[-\frac{1}{2}{\tilde g}^{\mu\nu}(\tilde{\nabla}_{\mu}\phi)\tilde{\nabla}_{\nu}\phi -V(\phi)\right] \nonumber
\\ &&+\int d^4 x\sqrt{-{\tilde g}}\, e^{-2\beta\kappa\phi} {\cal L}_{\rm m}[\chi_i,e^{-\beta\kappa\phi}{\tilde g}_{\mu\nu}]\ ,
\label{einsteinaction}
\ea
where the potential $V(\phi)$ is determined entirely by the original form~(\ref{jordanaction}) of the action and is given by
\be
V(\phi)=\frac{1}{2\kappa^{2}}\frac{R f_{R} - f}{(1+f_{R})^{2}} \ .
\label{einsteinpotential}
\ee
The Einstein-frame line element can be written in familiar Friedmann-Robertson-Walker (FRW) form as
\be
d{\tilde s}^2 ={\tilde a}^2(-d{\tau}^2+d{\bf x}^2) \ ,
\label{einsteinFRWmetric} 
\ee
where the Jordan and Einstein metrics are related through the conformal transformation ${\tilde a}^2\equiv e^{\beta\kappa\phi} \,a^2$. It is 
also convenient to define an Einstein-frame matter energy-momentum tensor by
\be
{\tilde T}_{\mu\nu} = ({\tilde \rho}+ {\tilde P}){\tilde U}_{\mu} {\tilde U}_{\nu} + {\tilde P} {\tilde g}_{\mu\nu}\ ,
\label{einsteinperfectfluid}
\ee
where ${\tilde U}_{\mu}\equiv e^{\beta\kappa\phi/2} \,U_{\mu}$, ${\tilde \rho}\equiv e^{-2 \beta\kappa\phi} \rho$ and ${\tilde P}\equiv e^{-2 \beta\kappa\phi} P$.

The equations of motion obtained by varying the action with respect to the metric ${\tilde g}_{\mu\nu}$ are
\be
{\tilde G}_{\mu\nu} = 8 \pi G \tilde{T}_{\mu \nu} + \frac{1}{2} \tilde{\nabla}_{\mu} \phi \tilde{\nabla}_{\nu} \phi + \frac{1}{2}(\tilde{g}^{\alpha \gamma} \tilde{\nabla}_{\alpha} \phi \tilde{\nabla}_{\gamma} \phi ) \tilde{g}_{\mu\nu} - V(\phi) \tilde{g}_{\mu\nu} \ ,
\label{einsteineom}
\ee
and are more familiar than those in the Jordan frame, although there are some crucial distinctions.  Most notably, in this frame test matter particles do not freely fall along geodesics of the metric ${\tilde g}_{\mu\nu}$, since the scalar field is also coupled to matter.  

The remaining equations of motion, for the scalar field and for the perfect fluid matter, are given respectively by
\bea
\phi'' + 2\tilde{\hub}\phi' + \tilde a^{2}V_{\phi} = \f{1}{2}\kappa\beta \tilde a^{2}(\tilde{\rho} - 3 \tilde{P} ) \ ,
\label{scalareom}
\\
\tilde\rho' +3\tilde{\hub}(\tilde{\rho} + \tilde{P}) = - \f{1}{2}\kappa\beta \phi'( \tilde\rho-3\tilde P) \ .
\eea
where  $V_{\phi}= dV/d\phi$.
\section{Perturbation Theory in the Jordan frame}
\label{perttheory}

In this section, we present the equations that govern the evolution of scalar perturbations in $f(R)$ theories in the Jordan frame.   We follow the notation of Kodama and Sasaki~\cite{Kodama} for general perturbations and that of Ma and Bertschinger~\cite{Ma:1994dv} in the conformal Newtonian and synchronous gauges. We use Latin and Greek letters to denote spatial and $4$-vector coordinates, respectively, and the Einstein summation convention is followed.

As in~\cite{Kodama}, we may write perturbations of 
the metric so as to separate the spatial and time dependences. For a given wave-number $k$ one can decompose the metric into four time dependent perturbations $A$, $B$, $H_{L}$ and $H_{T}$
\bea
\label{metric}
g_{00}&=&-a^{2}(1+2AY),\nonumber\\
g_{0i}&=&-a^2 BY_i, \nonumber\\
g_{ij}&=&a^2(\gamma_{ij}+2H_LY\gamma_{ij}+2H_TY_{ij}) \ ,
\label{en-mom-tensor}
\eea
where $\gamma_{ij}$ is the spatial metric and $Y=Y({k},{x})$ is the complete set of scalar harmonic functions. Here we consider the flat case, for which $Y\propto {\rm exp}(ik \cdot x)$. The perturbations in the energy momentum tensor are decomposed into 4 components: density, $\delta\rho\equiv \rho\delta$; velocity, $v$; isotropic pressure $\delta P$; and anisotropic stress $\f{3}{2}(\rho+P)\sigma$, via
\bea
{T^0}_0&=&-\rho[1+ \delta\, Y],\nonumber\\
{T^0}_j&=&(\rho+P)(v-B)Y_j,\nonumber\\
{T^i}_j&=&[P{\delta^i}_j + \delta P{\delta^i}_j Y +\f{3}{2}(\rho+P)\sigma{Y^i}_j] \ ,
\eea
where we use the notation of Ma and Bertschinger ~\cite{Ma:1994dv}  for the anisotropic stress (Kodama and Sasaki~\cite{Kodama} use the anisotropic stress perturbation $\Pi$, where $P\Pi = \f{3}{2}(\rho+P)\sigma$).

In  the Jordan frame, matter is minimally coupled and follows the geodesic of the usual metric $g_{\mu\nu}$. Thus, there is no dependence of the matter Lagrangian ${\cal L}_{\rm m}[\chi_i,g_{\mu\nu}]$ on our $f(R)$ modifications, and the conservation equations for matter do not differ from the conservation equations of standard general relativity,
\ba
\label{mattereom-generic1}
&&\delta'+(1+w)(kv+3H_L')+3\hub\l(\f{\delta P}{\delta \rho}-w\r)\delta=0\\
&&(v'-B')+\hub(1-3w)(v-B)+\f{w'}{1+w}(v-B)-\f{\delta P/\delta\rho}{1+w}k^2\delta-k^2A+\f{2}{3}k^2\sigma=0
\label{mattereom-generic2}
\ea

Two additional independent equations come from the four perturbed field equations. Defining $\delta R / Y$ by the following expression
\bea
&&\frac{\delta R}{Y}=\f{2}{a^{2}}\l[-6\frac{a''}{a}A-3\hub A' +k^{2}A +kB'+3k\hub B +9\hub H_{L}' +3 H_{L}''+2k^{2}\l(H_{L}+\frac{H_{T}}{3}\r) \r] \ ,
\eea
we provide all four equations here for completeness, and provide some intermediate results in appendix \ref{pertapp}.
\\
\noindent \underline{$0-0$ component}
\bea
&&(1+f_R)\l[6\hub^2A -2k\hub B -6\hub H_{L}' -2k^{2}\l(H_{L}+\frac{H_{T}}{3}\r)\r] +3f_{RR}\hub^{\prime}\, \f{\delta R}{Y} \nonumber
\\ &&  -\l(k^{2}f_{RR}+ 3\hub f_{RR}'\r)\f{\delta R}{Y} - 3\hub f_{RR}\l(\f{\delta R}{Y}\r)'+f_{R}'\l(6\hub A-kB-3H_{L}'\r) =-\kappa^{2} a^{2}\rho\delta  \ \ \ \ \ \ \ \ \
\eea


\noindent \underline{$0-i$ component}
\be
\l(1+f_R\r)\l[k\hub A-k\l(H_L+\f{H_T}{3}\r)^{\prime}\r]-\f{1}{2}k\l(f_{RR}\f{\delta R}{Y}\r)^{\prime}+\f{1}{2}k\hub\l(f_{RR}\f{\delta R}{Y}\r)+\f{1}{2}kf_R^{\prime}A=\f{\kappa^2a^2}{2}\l(\rho+P\r)\l(v-B\r)
\ \ \ \ \ \ \
 \ee
  

\noindent \underline{$i-i$ component}
\bea
&&2\l(1+f_R\r)\l[A\l(\hub^2+2\hub^{\prime}\r)-\f{1}{3}k^2A-\f{1}{3}k\l(B^{\prime}+\hub B\r)-\f{k}{3}\hub B-H_L^{\prime\prime}-2\hub H_L^{\prime}+\hub A^{\prime}-\f{1}{3}k^2\l(H_L+\f{H_T}{3}\r)\r] \nonumber
\\ &&+ \l(2\hub^2+\hub^{\prime}\r)f_{RR}\f{\delta R}{Y} -f_{RR}\l(\f{\delta R}{Y}\r)^{\prime\prime}-\hub f_{RR}\l(\f{\delta R}{Y}\r)^{\prime}-\f{2}{3}k^2f_{
RR}\f{\delta R}{Y}+2f_R^{\prime\prime}A\nonumber
\\ && +f_R^{\prime}\l(A^{\prime}+2\hub A-\f{2}{3}kB-2H_L^{\prime}\r)= \kappa^2a^2\delta P\ \ \ \ \ \ \ \ \
\eea


\noindent \underline{$i-j$ ($i\neq j$) component}
\bea
&& (1+f_{R})\l[-k^{2}A-k\l(B'+\hub B\r)+H''_{T}+\hub H'_{T}-k^{2}\l(H_{L}+\f{H_{T}}{3}\r)+\hub(H'_{T}-kB)\r] \nonumber
\\ && -k^2f_{RR}\frac{\delta R}{Y}-f_R^{\prime}\l(kB-H_T^{\prime}\r)=\kappa^{2} a^{2}\f{3}{2}(\rho+P)\sigma 
\eea

\subsection{Conformal Newtonian Gauge}

In the conformal Newtonian gauge $H_{T}=B=0$, $A=\psi$ and $H_{L} = -\phi$, where we have used the notation of Ma and Bertschinger (Kodama and Sasaki use $H_{L} = \Phi$ and $H_{T}=\Psi$). In this gauge, the evolution equations (\ref{mattereom-generic1}) and (\ref{mattereom-generic2}) for a cold dark matter (CDM) and radiation overdensity may be combined in single second order differential equations,
\ba\label{matter-confNew}
\delta_c '' +\hub \delta_c' + k^{2}\psi -3\phi '' -3\hub \phi' &=& 0
\\
\delta_{\gamma}''+\f{1}{3}k^2\delta_{\gamma}+\f{4}{3}k^2\Psi-4\Phi''=0
\ea
These, in combination with the following two independent equations, are sufficient to fully specify the evolution 
\\
\noindent 
\underline{$i-j$ ($i\neq j$) component}
\be\label{einstein-ij-conNew}
(1+f_R) \left( \psi - \phi \right) + f_{RR} \dRoY = -\f{a^{2}}{k^2}\kappa^2\f{3}{2}\Sigma_i (\rho_i+P_i)\sigma_i
\ee

\noindent \underline{$0-0$ component}
\bea\label{Einstein-00-conNew}
&& (1+f_R)\l[ 2k^{2}\phi +6\hub\l( \phi'+\hub\psi  \r)\r]
  +3f_{RR} \hub^{\prime}\,\f{\delta R}{Y} + \nonumber
  \\ && -\l(k^{2}f_{RR}+  3\hub f_{RR}'\r)\f{\delta R}{Y}  - 3\hub f_{RR}\l(\f{\delta R}{Y}\r)'+f_{R}'\l(6\hub \psi+3\phi'\r)=- \kappa^2 a^{2}\Sigma_i \rho_i\delta_i  \ ,
\eea
where the perturbed expression for the Ricci scalar is now
\bea
\frac{\delta R}{Y}&=&\f{2}{a^{2}}\l[-6\frac{a''}{a} \psi-3\hub \psi' + k^{2}\psi-9\hub \phi' -3\phi''-2k^{2}\phi \r]. \   \ \ \ \ \ \ \
\eea

\subsection{Synchronous gauge}

In the sychronous gauge $A=B=0$ and, following Ma and Bertschinger, $H_{L} = h/6 $ and $H_{T} = -3(\eta+h/6)$. To completely define the synchronous coordinates we may remove the remaining freedom by specifying that cold dark matter particles have zero peculiar velocity $v$ in this gauge. The evolution equations for cold dark matter and radiation then reduce to, 
\begin{eqnarray}\label{mattereom-generic-full-syn}
&& \delta_c' =-\f{1}{2}h' \\
&& \delta_\gamma '' + \frac{k^{2}}{3}\delta_\gamma-\frac{4}{3}\delta_{c}'' =0.
\end{eqnarray}

In this gauge it is again possible to use the following two independent equations to fully specify the evolution 

\noindent \underline{$i-j$ ($i\neq j$) component}

  \be\label{einstein-ij-synchronous)}
 (1+f_{R})\l[2k^2\eta-\l(h^{\prime\prime}+6\eta^{\prime\prime}\r)-2\hub\l(h^{\prime}+6\eta^{\prime}\r)\r]-f_R^{\prime}\l(h^{\prime}+6\eta^{\prime}\r)- k^{2}f_{RR}\f{\delta R}{Y} =\kappa^{2} a^{2}3\Sigma_i(\rho_i+P_i)\sigma \ \ \ \ \ \ \ \ 
 \ee

\noindent \underline{$0-0$ component}
\bea\label{Einstein-00-Synch}
&& \l(1+f_R\r)\l[-\hub h^{\prime}+2k^2\eta\r]+3f_{RR}\hub^{\prime}\,\f{\delta R}{Y}  -\l(k^{2}f_{RR}+  3\hub f_{RR}'\r)\f{\delta R}{Y}  - 3\hub f_{RR}\l(\f{\delta R}{Y}\r)'-\frac{1}{2}f_{R}' h^{\prime}=- \kappa^2 a^{2}\Sigma_i\rho_i\delta_i \ ,
\eea
with the perturbed Ricci Scalar now given by
\bea
\frac{\delta R}{Y}&=&\f{2}{a^{2}}\l[\f{3}{2}\hub h' +\f{1}{2} h''-2k^{2}\eta \r] \ .
\eea

\section{Structure formation in late-time dominating $f(R)$ theories}
\label{evol}

We now consider in more detail, the solutions to perturbation evolution in scenarios in which the $f(R)$ coupling becomes important at late times. We first consider evolution in the Einstein frame, since it is intuitively somewhat simpler, for pressureless matter, radiation and a scalar field, neglecting the interaction between baryons and radiation and then transform results to the Jordan frame using the approach outlined in appendix \ref{mappingpert}.

As in section \ref{conformal}, the Einstein frame variables (except the scalar field) are denoted by a tilde. In the Einstein frame we have background equations
\bea
\tilde\hub^{2} &=& \frac{\kappa^{2}}{3}\left(\f{\phi'^{2}}{2}+\tilde a^{2} V(\phi)+\tilde a^{2} \tilde\rho_{c}+ \tilde a^{2} \tilde\rho_{\gamma} \right)
\\
\phi''&+&2 \tilde\hub\phi' + \tilde a^{2} V_{\phi}= \frac{1}{2}\kappa\beta \tilde a^{2} \tilde\rho_{c}
\\
\tilde\rho_{c} &\equiv &\tilde\rho_{c}^{*}\exp\l(-\frac{\kappa\beta\phi}{2}\r)
\\
\tilde\rho_{c}^{*} &\equiv & \frac{\tilde\rho_{c}^{*0}}{\tilde a^{3}}
\eea
where $\tilde\rho_{c}^{*0}$ is a constant and the scalar field is decomposed into a time dependent background and spatial varying perturbation $ \phi(t) +\delta\phi(x,t)$. As discussed in \cite{Bean:2001ys}, a scalar coupling to dark matter does not change the first order perturbation equations for CDM if we define the perturbation with respect to $\tilde\rho_{c}^{*}$
\bea
\tilde\delta_{c}&\equiv&\frac{\delta \tilde\rho_{c}^{*}}{\tilde\rho_{c}^{*}}
\\
\tilde\theta_{c}&\equiv& ik^{j}v_{j}^{*} \ ,
\eea 
where $\tilde\theta_{c}$ is defined consistently with $\tilde\delta_{c}$.
With the residual freedom in the synchronous gauge we can then set $\tilde\theta_{c} =0 $ which relates $\tilde\delta_{c}=-\frac{1}{2}\tilde h$ as in GR. The perturbation in the Jordan frame is then given by
\bea
\delta_{c} 
&=&\tilde\delta_{c}+\frac{3}{2}\kappa\beta\delta\phi
\eea
The perturbation equations in the synchronous gauge are subsequently
\bea
\tilde\delta_{c}'' + \tilde\hub \tilde\delta_{c}' -\frac{3}{2} \tilde\hub^{2}(2 \tilde\Omega_\gamma \tilde \delta _\gamma + \tilde\Omega_{c}(\tilde\delta_c-\frac{1}{2} \kappa\beta\delta\phi) + 2\kappa^{2}\phi'\delta\phi'-\kappa^{2}V_{\phi}\delta\phi 
&=&0 \label{CDMeq}
\\
\delta\phi '' +2 \tilde\hub \delta\phi' +k^{2}\delta\phi + \tilde a^{2}V,_{\phi\phi}\delta\phi - \phi' \tilde\delta_{c}'-  \frac{3 \beta}{2\kappa} \tilde\hub^{2} \tilde\Omega_{c}(\tilde\delta_c-\frac{1}{2}  \kappa\beta\delta\phi)&=&0
\\
\tilde\delta_{\gamma}''+\frac{1}{3}k^{2} \tilde\delta_{\gamma}-\frac{4}{3} \tilde\delta_{c}'' &=& 0.
\eea

\underline{{\it Radiation dominated era:}} In the radiation era, $\tilde a\propto \tau$, the coupling is negligible and the perturbation evolution equations become
\bea
\tilde\delta_{c}'' + \frac{1}{\tau} \tilde \delta_{c}' -\frac{3}{\tau^{2}} \tilde \delta _\gamma & \simeq& 0
\\
 \delta\phi '' +\frac{2}{\tau} \delta\phi'-   \frac{3 \beta}{2\kappa}\frac{1}{\tau^{2}
} \tilde\Omega_{c} \tilde\delta_c& \simeq&0
\\
\tilde\delta_{\gamma}''+\frac{1}{3}k^{2} \tilde\delta_{\gamma}-\frac{4}{3} \tilde\delta_{c}''& \simeq& 0
\eea
which yields solutions, fixing $\tilde\delta_{ci}\equiv \tilde\delta_{c}(\tau_{i})$
\bea
\tilde\delta_{c} &\simeq& \tilde\delta_{c i}\l(\frac{\tau}{\tau_{i}}\right)^{2} \\
\tilde\delta_{\gamma}& \simeq& \frac{4}{3}\frac{1}{[1+\frac{(k \tau)^{2}}{6}]} \tilde\delta_{c}. \eea
For the scalar field the dominant driver is the CDM density fluctuation, and yields,
\bea
\frac{2}{\tau}\delta\phi' +k^{2}\delta\phi &\approx &\frac{3 \beta}{2 \kappa} \frac{1}{\tau^{2}
} \tilde\Omega_{c} \tilde\delta_c
\\ 
& \simeq& \frac{3 \beta}{2 \kappa} \frac{1}{\tau^{2}
}\frac{ \tau}{\tau_{eq}} \delta_{c i}\tau^{2}    \ \ \ \ \ \ \ \ \ ( \tilde\rho_{\gamma}\gg \tilde\rho_{c})
\\ 
\delta\phi& \simeq& \frac{\beta}{4 \kappa}\frac{\tilde\rho_{c}^{i}}{\tilde\rho_{\gamma}^{i}} \frac{1}{[1+\frac{(k\tau)^{2}}{6}]}
 \tilde\delta_{c i} \l(\frac{\tau}{\tau_{i}}\right)^{3}
\eea
In the radiation era, the coupling is negligible and the Einstein and Jordan frames are comparable. In terms of the Jordan frame expansion rate, $a$, we find
\bea
\tilde\delta_{c}& \simeq&  \tilde\delta_{c i}\l(\frac{a}{a_{i}}\r)^{2}
\\
\frac{\tilde\delta_{\gamma}}{\tilde\delta_{c}} &\simeq& \frac{4}{3}\frac{1}{[1+\frac{(k \tau)^{2}}{6}]}  
\\
\frac{\kappa\delta\phi}{\beta\delta_{c}}& \simeq&\frac{1}{4}\frac{\tilde\rho_{c}^{i}}{\tilde\rho_{\gamma}^{i}}\frac{1}{[1+\frac{(k \tau)^{2}}{6}]}
\l(\frac{a}{a_{i}}\r)
\eea
so that the scalar field perturbation evolves more rapidly than the matter and radiation density contrasts but has a smaller initial value. The amplitude of the scalar field perturbation is $k$ dependent, with larger wavenumbers being comparatively suppressed.

At the matter radiation transition the scalar field density fluctuation is approximately
\bea
\frac{\kappa\delta\phi_{eq}}{\beta\tilde\delta_{c,eq}} & \simeq& \frac{1}{4}\frac{1}{[1+\frac{(k\tau)^{2}}{6}]}\label{eq1}
\eea
Because of the suppression of the scalar field fluctuations, the Jordan CDM density fluctuations in the radiation era is comparable to the Einstein CDM density fluctuation
\bea
\delta_{c}(a) &=& \tilde\delta_{c} +\frac{3}{2}\kappa\beta\delta\phi
\\
&\simeq &    \delta_{c i}\l(\frac{a}{a_{i}}\r)^{2}
\eea

\underline{{\it Matter dominated era:}}  In $f(R)$ models the background evolution approaches an attractor solution in the matter dominated era in which, $\tilde a\propto \tilde t^{\frac{3}{5}}$ ($\tilde a\propto \tau^{\frac{3}{2}}$), which in the Jordan frame corresponds to  $a\propto {t}^{\frac{1}{2}}$  ($a\propto \tau$)\cite{Amendola:2006mr}. In appendix \ref{scaling} we highlight how this result is obtained. The scaling regime has a purely kinetic scalar field with $\tilde\Omega_{\phi}=1/9$, and $\tilde\Omega_{c}=8/9$, leading to 
\bea
\phi' &=& \frac{1}{\beta\kappa}\frac{1}{\tau}
\eea
The matter and scalar field equations are,
\bea
\tilde\delta_{c}'' + \frac{3}{2}\frac{1}{\tau} \tilde\delta_{c}' -3\frac{1}{\tau^{2}}(\tilde\delta_c-\frac{1}{2} \kappa\beta\delta\phi) +  \frac{2\kappa}{\beta}\frac{1}{\tau}\delta\phi'
&=&0
\\
\delta\phi '' +2 \tilde\hub \delta\phi' +k^{2}\delta\phi -  \frac{1}{\beta\kappa}\frac{1}{\tau} \tilde\delta_{c}'-  \frac{3\beta}{\kappa} \frac{1}{\tau^{2}} (\tilde\delta_c-\frac{1}{2}  \kappa\beta\delta\phi)&=&0
\eea
Denoting $x =\kappa\delta\phi_{eq}/\beta\tilde\delta_{c,eq}$, for modes $k\tau_{eq}>1$ the scalar field is always going to be negligible to the matter perturbation. Likewise for scales for which $1/\tau_{0}< k< 1/\tau_{eq}$ suppression of the scalar field will set in during the matter dominated era. For these instances, where the scalar field is negligible, we find a scaling solution
\bea
\tilde\delta_{c} (\tau)&=&\tilde\delta_{c,eq}\left(\frac{\tau}{\tau_{eq}}\right)^{\frac{3}{5}} 
\\ &=& \tilde\delta_{c,eq}\left(\frac{a}{a_{eq}}\right)^{\frac{3}{2}} 
\eea
For modes for  which $k\tau_{eq}<1$ i.e. $x_{eq}\sim 1/4$ the scalar field approaches equivalence with the dark matter perturbations in the matter dominated era. We find a scaling solution of the form
\bea
x=\frac{\kappa\delta\phi}{\beta\tilde\delta_{c}} &=& \frac{21}{25}\frac{1}{1+\frac{4}{25}(k\tau)^{2}}\label{eq2}
\eea
For scales with $k\tau_{eq}>1$ the scalar field perturbation is suppressed in the matter era and $\delta_{c}\simeq \tilde\delta_{c}$. For modes $k\tau_{eq}<1$, the scalar field scaling solution (\ref{eq2}) leads to the Jordan frame perturbation being slightly boosted above that of the Einstein frame, 
\bea
\delta_c = \l(1+\frac{14}{25}\frac{1}{1+\frac{4}{25}(k\tau)^{2}}\r)\tilde\delta_{c} 
\eea
this occurs until $\tau \simeq 1/k$ at which point the scalar field becomes suppressed and $\delta_{c}\simeq \tilde\delta_{c}$.

\underline{{\it Accelerating era:}} 
During the accelerating era, the Einstein frame CDM density decreases to effectively zero, so that if $\tilde a \propto \tau^{2/(1+3w_{eff})}$, where $w_{eff}$ is the effective Einstein frame equation of state
\bea
\tilde\delta_{c}'' + \frac{2}{(1+3\tilde{w}_{eff})}\frac{1}{\tau} \tilde\delta_{c}' & \approx & 0
\\
\delta\phi '' +2 \tilde\hub \delta\phi'+\tilde a^{2}V,_{\phi\phi}\delta\phi &\approx& 0
\eea
The potential obtained in the Einstein frame typically tends towards an exponential potential at large $\phi$, $V\sim\exp{-\lambda\kappa\phi}$. As discussed in the appendix, the accelerating regime is then an attractor in which $w_{eff} = -1+\lambda^{2}/3$, and the scalar field perturbation evolves as $\delta\phi\propto a ^{p}$ with
\bea
p &=&- \frac{6-\lambda^{2}}{2-\beta\lambda}\left[1\pm \sqrt{1-8/\left(\frac{6}{\lambda^{2}}-1\right)}\right]
\eea
For $\lambda<\sqrt{6}$, $p<0$ with a complex amplitude and the scalar perturbations decay yielding $\tilde\delta_{c}\simeq \delta_{c}=$ constant.  As a result, in the Jordan frame, $\delta\rho_{c}=\rho_{c}\delta_{c}$ decays $\sim a^{-3}$. In the current era the classical perturbations are still observed; however the decay implies that, as the accelerating era continues, the classical perturbations will be smoothed, and quantum fluctuations generated in the era will ultimately become important, as the cosmic no hair theorem requires, just as in early universe inflation \cite{Mukhanov:1990me}.

It is common in $f(R)$ theories that the transition from scaling behavior to acceleration occurs significantly earlier than in the $\Lambda$CDM scenario. This leads to the late time suppression of the density fluctuations  being more pronounced with important implications for comparison with observational data. We discuss the evolution and implications of structure formation observations for two specific examples below.

\subsection{Examples: $f(R)= -\mu^{4}/R$ and $f(R)= -\mu_{1}H_{0}^{2}\exp(-R/\mu_{2}H_{0}^{2})$}

It is important to note that the results above are true for any form of $f(R)$ where the coupling dictates the expansion rate during the matter dominated era. Such behavior is not sensitive to the details of the potential and is expected in a large class of $f(R)$, including all explicit functional forms proposed so far in the literature.

 Here, for concreteness, however we provide numerical results for two specific examples,  $f(R)= -\mu^{4}/R$ and $f(R)= -\mu_{1}H_{0}^{2}\exp(-R/\mu_{2}H_{0}^{2})$, with $\Omega_{m}^{eff} = \kappa^{2}\rho_{m}(a=1) / 3\hub^{2} = 0.3$ and $H_{0}=70/kms^{-1}Mpc^{-1}$. 
  
\begin{figure}[h]
\begin{center}
\hbox{\includegraphics[width=90mm]{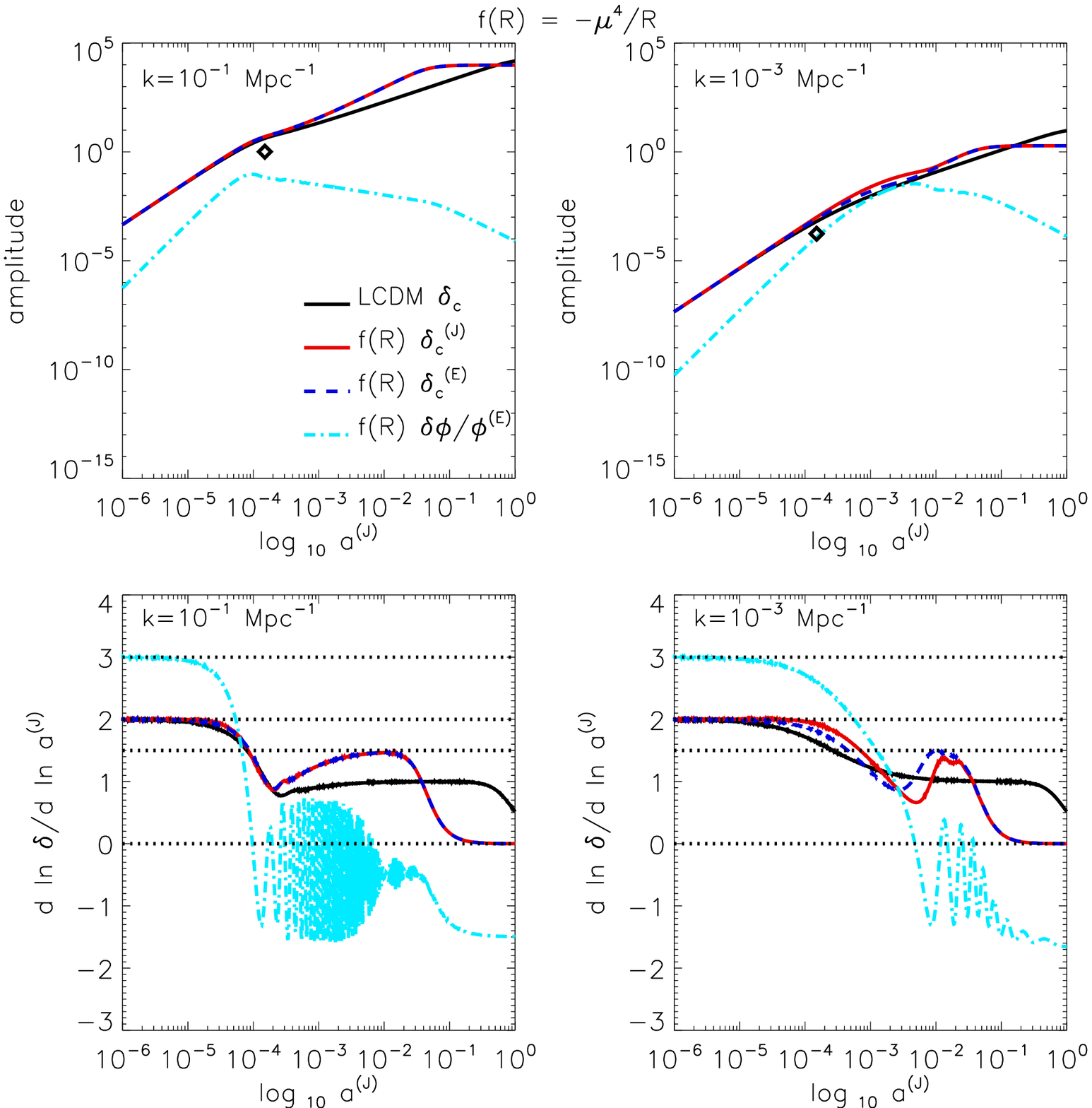}\includegraphics[width=90mm]{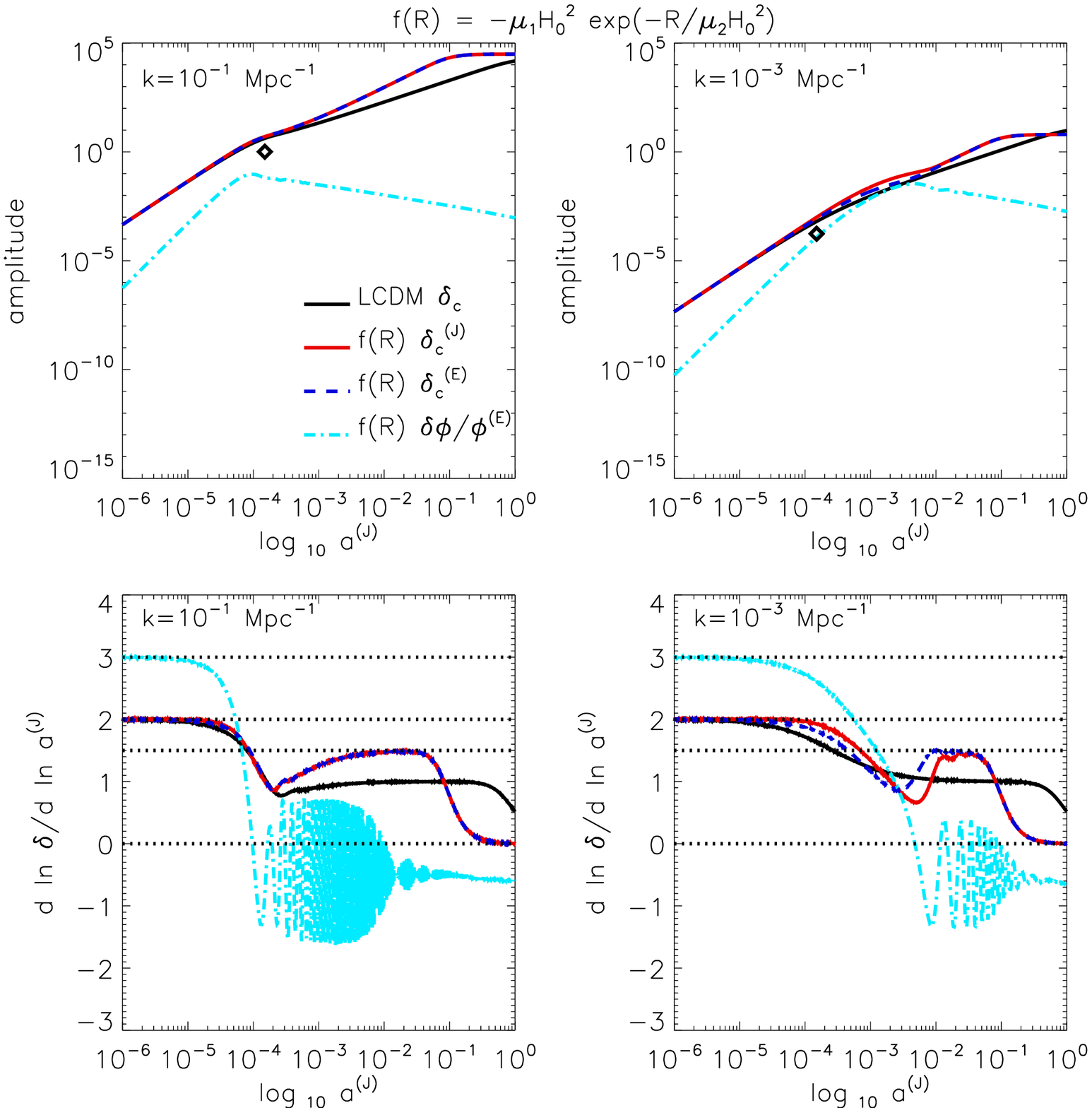}
}\end{center}
\caption{Perturbation evolution for [left panel] the $f(R)=- \mu^{4}/R$  and [right panel] the $f(R)=- \mu_{1}H_{0}^{2}\exp(-R/\mu_{2}H_{0}^{2})$ model with $\Omega_{m}^{eff} = \rho_{m}^{0}/3H_{0}^{2} =0.3$ and $\mu^{4}$, and $\{ \mu_{1}, \mu_{2}\}$ chosen in each case to give $H_{0} = 70 kms^{-1}Mpc^{-1}$. [Top panels] The density fluctuations for the f(R) theory are compared to that for the equivalent $\Lambda$CDM scenario for two comoving scales $k=10^{-1}Mpc^{-1}$ (left) and  $k=10^{-3}Mpc^{-1} $ (right). The diamond shows the analytic value of $x_{eq}$ in the limit of no suppression from $k\tau_{eq}>1$, as described in (\ref{eq1}). [Lower panels] The power law evolution of the density fluctuations $d\delta /d\ln a$ for the different density components. The main scaling solutions are shown by dotted lines, respectively  $d\ln\delta /d\ln a=3$ and $d\ln\delta /d\ln a=2$ for the scalar field  and matter components in the radiation era and $d\ln\delta /d\ln a=1.5$ and $d\ln\delta /d\ln a=0$ for the matter perturbations in the matter dominated and accelerated eras. In the figure $(E)$ and $(J)$ denote the Einstein and Jordan frame quantities respectively.}
\label{fig:fofR}
\end{figure}

 In figure \ref{fig:fofR}, we show the evolution for the specific examples in comparison to an equivalent $\Lambda$CDM scenario for two different comoving scales $k=10^{-3}Mpc^{-1}$ and $k= 10^{-1}Mpc^{-1}$ relevant to galaxy structure and CMB observations respectively.  One can see the analytical scaling solutions derived above hold in the radiation and matter dominated eras. 

\begin{figure}[h]
\begin{center}
\hbox{\includegraphics[width=90mm]{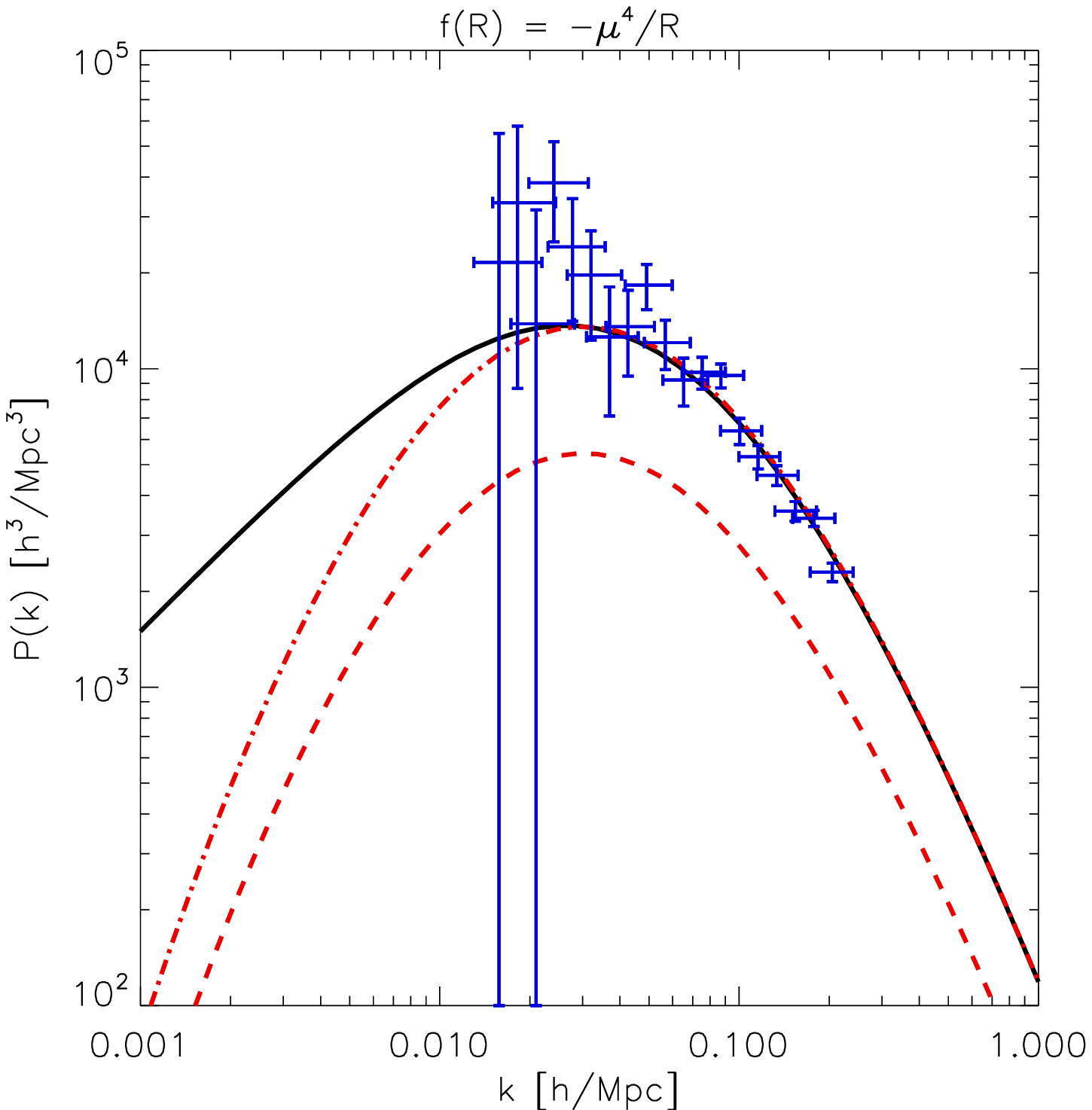}\includegraphics[width=90mm]{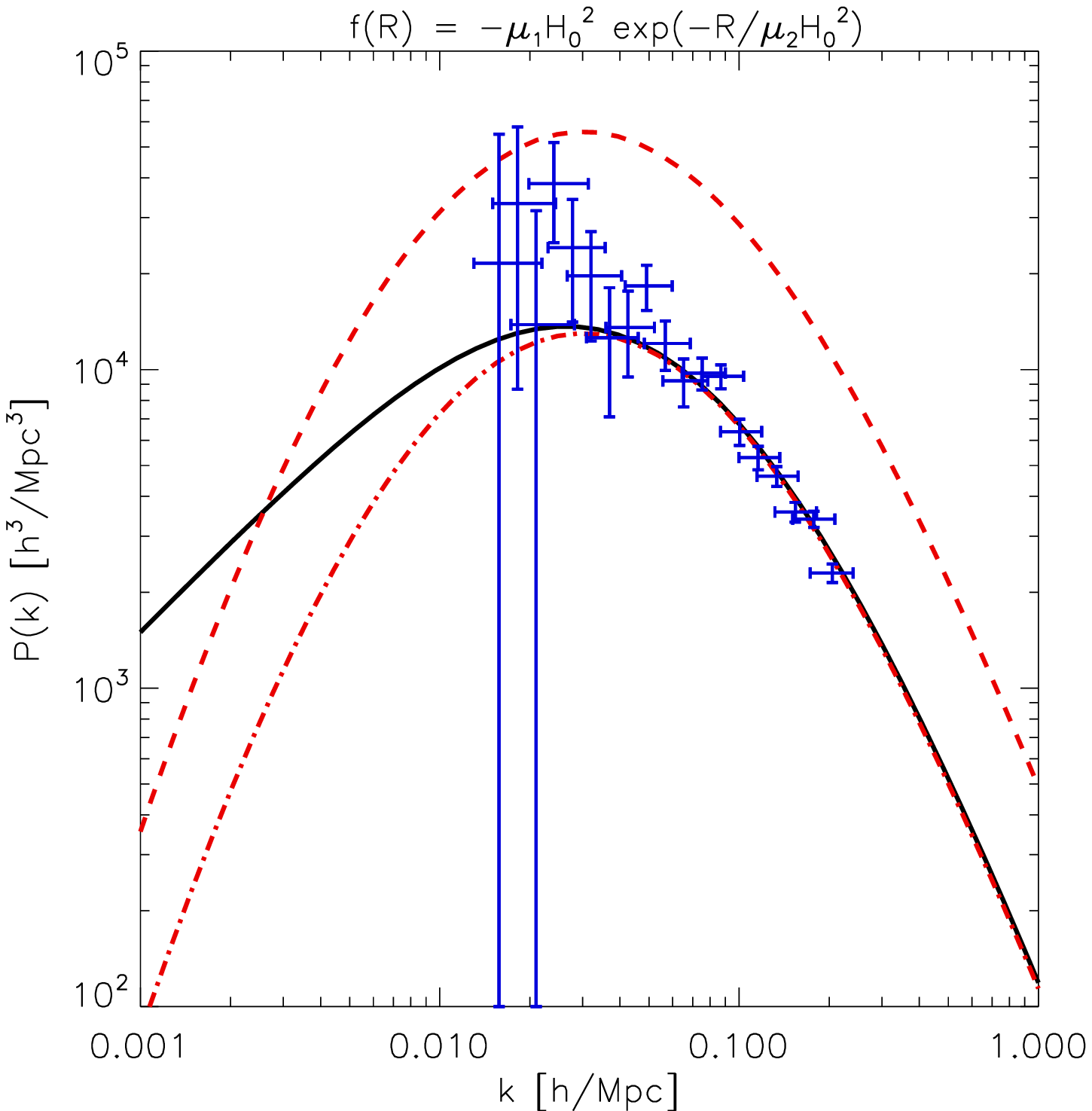}
}\end{center}
\caption{The matter power spectrum for $\Lambda$CDM (full black) and  [left panel] the $f(R)= -\mu^{4}/R$ model  and [right panel] the $f(R)= -\mu_{1}H_{0}^{2}\exp(-R/\mu_{2}H_{0}^{2})$ model for the same normalization (red dashed) and for normalization to give small scale agreement between the two models (red dot-dashed) are shown against the SDSS matter power spectrum data \cite{Tegmark:2003uf}. One can see that the f(R) model cannot simultaneously give small scale agreement  with galaxy matter power spectrum and large scale agreement with the CMB.}
\label{fig:pk}
\end{figure}

The early onset of acceleration  in comparison to the $\Lambda$CDM scenario leads to  increased relative  suppression of the large scale density fluctuations and inconsistencies with the galaxy matter power spectrum and ISW effect \cite{Zhang:2005vt,Song:2006ej}. In figure \ref{fig:pk} we show the effect on the matter power spectrum at $z=0$. For the same primordial normalization  the late time evolution leads to an overall large scale suppression of the matter power spectrum in comparison to the $\Lambda$CDM scenario. If one arbitrarily renormalizes the spectrum to be in agreement with galaxy matter spectra, then the spectrum still shows too great a suppression at large scales to be consistent with the CMB.

\subsection{Comment on the subhorizon CDM over density in the Jordan frame}

  In ~\cite{Zhang:2005vt} a potential problem was observed in the conformal Newtonian gauge in the Jordan frame, in that there was an apparent zeroing of the CDM matter density fluctuation at critical scales. This is obviously at odds with the results of the previous section so we revisit it here to highlight the origins of, and explain, the discrepancy.

As seen in section \ref{perttheory} evolving the perturbation equations in the Jordan frame is significantly more laborious than in the Einstein frame. In order to simplify the equations to study the evolution, therefore, in \cite{Zhang:2005vt}  made a {\it quasi-static} approximation at sub-horizion scales where $k/\hub\gg1$ . In this approximation, for each perturbation variable, one assumes that the time derivative is negligible in comparison to its spatial derivative.  

In the subhorizon limit, the matter evolution equation~(\ref{matter-confNew}) would simplify to
\bea\label{2nd-delta-prep}
\delta_{c}''+\hub \delta_{c}' + k^{2}\psi = 0 \ .
\eea
If, following \cite{Zhang:2005vt}, we define the parameter $Q$ as
\begin{equation}\label{defQ}
Q \equiv -2 \koa{2} \frac{f_{RR}}{1+f_R} \ ,
\end{equation}
the independent components of the Einstein equations~(\ref{einstein-ij-conNew}) and~(\ref{Einstein-00-conNew}) become
\bea
\delta_{c}''+\hub\delta'_{c}  -\frac{1}{2(1+f_{R})}
\kappa^{2}a^{2}\l(2\rho_\gamma \delta _\gamma + \frac{(1-2Q)}{\l(1-\frac{3}{2}Q\r)} \rho_c\delta_c\r) 
\approx 0\label{delta-evolution}
\eea
where note that $\kappa^{2}a^{2}(\rho_{\gamma}+\rho_{c})\neq 3\hub^{2}$ here. Compare this to the corresponding scale-independent behavior in standard general relativity
\begin{equation}\label{delta-late-stdGR}
\delta_{c} ''+  \hub \delta_{c}'- \frac{1}{2}\kappa^{2}a^{2}(2\rho_\gamma \delta _\gamma + \rho_{c} \delta_{c} ) = 0 \ .
\end{equation}

Inspection of the overdensity evolution equation~(\ref{delta-evolution}) leads to an identification of a critical value $Q = 2/3$, at which the overdensity $\delta_{c}$ is driven to zero  with two additional apparent zero density points at $Q = 1/2$ and $Q = 1$, in the spatial off-diagonal Einstein equations. This apparent zeroing of the CDM overdensity also arises, under the same assumptions, in the synchronous gauge.

However no such zeroing of $\delta_{c}$ is seen to arise from the evolution and subsequent conformal transformation of the full, unapproximated equations in the Einstein frame.
The discrepancy arises because the quasi-static approximation used is too aggressive and removes important information about the evolution. We can see this from doing a conformal transformation of the Einstein frame CDM perturbation equation (\ref{CDMeq}), for which we obtain,
\bea
&&\ddot{\delta}_{c}+ \hub\dot{\delta}_{c}-\frac{1}{2}\kappa^{2}a^{2}(2\rho{\gamma}\delta_{\gamma}+\rho_{c}\delta_{c})
- \frac{3}{2} \frac{d^{2}}{d\tau^{2}}\l(\frru\delta R\r) -\kappa^{2}a^{2}(\rho_{\gamma}+\rho_{c})\frru\delta R
  \nonumber \\
&&+3\frac{\dot{f}_{R}}{(1+f_{R})}\frac{d}{d\tau}\l(\frru\delta R\r)
 -4\frac{f_{RR}}{(1+f_{R})}\l[ \frac{Rf_{R}-f}{(1+f_{R})^{2}}\r]\delta R=0
\eea
   The quasi-approximation therefore, in which one neglects time derivatives for $f_{RR}$,  used to infer the zeroing in the Jordan frame, is invalid, since zero, first and second time derivatives of $f_{RR}/(1+f_{R})$ all come into play in the conformally transformed full equation.

\section{Conclusions}
\label{conclusions}

The possibility that cosmic acceleration is our first signal of a far-infrared modification of 
General Relativity is a logical alternative to the cosmological constant and to dark energy models.

In this paper, we have focused on $f(R)$ modifications to GR, exploring the details of 
cosmological perturbation theory and its implications for the linear theory of structure formation on the universe. The details of such an analysis are complicated by the modified dynamics and we find that it is simpler to conduct the analysis in the Einstein frame. In particular, we have seen that the use of the `quasi-static' approximation in the Jordan frame can lead to misleading, incorrect conclusions, suggesting that the overdensity is driven to zero at specific scales. This is inconsistent with the conclusions obtained when one uses the conformally transformed results from the full equations in the Einstein frame.

We have shown, considering the evolution in the Einstein and Jordan frames, and general and specific choices of gauges, that the evolution of density fluctuations in $f(R)$ gravity leads to predictions that are inconsistent with cosmological observations. The large scale density fluctuations are suppressed in comparison to small scales, leading to an inability to fit both small scale galaxy data and large scale, CMB, data simultaneously. These findings hold for the wide class of $f(R)$ that behave like GR at early times and only diverge from GR at low curvatures, in the matter dominated era. As such, these results rule out a wide class of modified gravity models aimed at providing an alternative explanation to the dark energy problem.

\acknowledgments

We thank Eanna Flanagan, Ira Wasserman and  Pengjie Zhang for helpful discussions in the course of this work. RB and DB's work is supported by National Science Foundation grants AST-0607018 and PHY-0555216 and an Affinito-Stewart grant from the President's Council of Cornell Women. The work of LP, AS and MT was supported in part by the National Science Foundation under grant NSF-PHY0354990, by funds from Syracuse University and by Research Corporation.

\appendix
\section{Jordan frame perturbation equations}\label{pertapp}

We summarize here some of the components that are used in the derivation of the results in section \ref{perttheory}. The perturbations to the geometric quantities are unmodified in $f(R)$ theories. For the Christoffel symbols we have:
\bea\label{Christoffel-perturbed}
&&\delta{\Gamma^0}_{00}=A^{\prime}Y,\hspace{0.2cm}\delta{\Gamma^0}_{0j}=-\left[kA+\hub B\right]Y_j,\\
&& \delta{\Gamma^0}_{ij}=\left[-2\hub A+(k/3)B+2\hub H_L+H_L^{\prime}\right]\gamma_{ij}Y+\left[-kB+2\hub H_T+H_T^{\prime}\right]Y_{ij}\\
&&\delta{\Gamma^j}_{00}=-\left[kA+B^{\prime}+\hub B\right]Y^j,\\
&&\delta{\Gamma^i}_{0j}=H_L^{\prime}{\delta^i}_jY+H_T^{\prime}{Y^i}_j\\
&&\delta{\Gamma^i}_{jk}=-kH_L\left({\delta^i}_jY_k+{\delta^i}_kY_j-\gamma_{jk}Y^i\right)+\hub B\gamma_{jk}Y^i+H_T\left({Y^i}_{j|k}+{Y^i}_{k|j}-{Y_{jk}}^{|i}\right].
\eea
And for the Ricci scalar and Ricci tensor,
\bea
&&\delta R=\f{2}{a^{2}}\l[-6\frac{a''}{a}A-3\hub A' +k^{2}A +kB'+3k\hub B+9\hub H_{L}' +3 H_{L}''+2k^{2}\l(H_{L}+\frac{H_{T}}{3}\r) \r] Y\\
&&\delta R_{00} =-\l[k^{2}A-3\hub A'+k(B'+\hub B)+3H_{L}''+3\hub H_{L}'\r]Y\\
&& \delta R_{kj} = \l[-2\l(\f{a''}{a}+\hub^2\r)A-\hub A' +\f{k^{2}}{3} A+\f{k}{3}\l(B'+\hub B\r)+\f{4}{3}\hub kB+\r.\nonumber
\\ &&\l.+H''_{L}+5\hub H'_{L}+2\l(\f{a''}{a}+\hub^2\r)H_{L}+\f{4k^{2}}{3}\l(H_{L}+\f{H_{T}}{3}\r)\r]\delta_{kj}Y\nonumber
\\ && 
+\l[-k^{2}A-k\l(B'+\hub B\r)+H''_{T}+\hub H'_{T}+ 2\l(\f{a''}{a}+\hub^{2}\r)H_{T}-k^{2}\l(H_{L}+\f{H_{T}}{3}\r)+\hub(H'_{T}-kB)\r] Y_{kj} 
\\&& \delta R_{0j} = \l[-\l(\f{a''}{a}+\hub^{2}\r)B-2k\hub A + 2kH_{L}'+\f{2}{3}kH_{T}'\r]Y_{j}
\eea
The main additional components that need to be calculated come from the perturbations to the covariant derivative terms in (\ref{jordaneom})
\bea
\delta(\nabla_{\mu}\nabla_{\nu} F) 
&=& \nabla_{\mu}\nabla_{\nu} \delta F - \delta \Gamma^{\beta}_{\mu\nu}\d_{\beta}F 
\\
\delta(\nabla^{\mu}\nabla_{\nu} F) &=&
\nabla^{\mu}\nabla_{\nu} \delta F +\delta g^{\mu\alpha} \nabla_{\alpha}\nabla_{\nu}F - g^{\mu\alpha} \delta \Gamma^{\beta}_{\alpha\nu}\d_{\beta}F  \ \ \ \ \ \ \ 
\eea
We are considering a function $F=f_R$ where $f_{R} = f_{R}(t) + \delta f_{R}(x,t)$,
\bea
\delta(\nabla^{i}\nabla_{i} f_{R})&=& \l[ -\f{k^{2}}{a^{2}}f_{RR}\f{\delta R}{Y} - \f{3}{a^{2}}\hub \l(f_{RR}\f{\delta R}{Y}\r)' +\f{f_{R}'}{a^{2}}\l(6\hub A-kB-3H_{L}'\r)\r]Y
\\
\delta(\nabla^{0}\nabla_{0} f_{R}) &=&\l[-\f{ \l(f_{RR}\f{\delta R}{Y}\r)''}{a^{2}} +\f{\hub}{a^2}\l(\f{f_{RR}\delta R}{Y}\r)^{\prime}+2\f{f_{R}''}{a^{2}}A-2A\hub f_R^{\prime}+\f{f_{R}'}{a^{2}}A'\r]Y 
\\
\delta(\nabla_{i}\nabla_{j} f_{R}) &=&\l[ -\f{k^{2}}{3}f_{RR}\f{\delta R}{Y} - \hub \l(f_{RR}\f{\delta R}{Y}\r)'\r.\nonumber 
 \l.+ f'_{R} \l(2\hub A-\f{k}{3}B-2\hub H_{L}-H_{L}'\r)\r]\delta_{ij}Y +\nonumber\\
 &+& \l[ k^{2}f_{RR}\f{\delta R}{Y}+f'_{R}\l(kB-2\hub H_{T}-H_{T}'\r)\r]Y_{ij} \nonumber
\\ && 
\\
\delta(\nabla_{0}\nabla_{j} f_{R})&=&\l[-kf_{RR}'\l(\f{\delta R}{Y}\r) - kf_{RR} \l(\f{\delta R}{Y}\r)' + k\hub f_{RR}\l(\f{\delta R}{Y}\r)+ f_{R}'\l(kA+\hub B\r) \r]Y_{j}
\eea

\section{Equations for the Mapping of Perturbation Theory to the Einstein frame}
\label{mappingpert}

As reviewed in section \ref{conformal} , there is a particular conformal transformation which maps the Jordan frame action into a Hilbert-Einstein action for gravity, with the  introduction of a scalar field $\phi(\vec{x},t)$. In this Einstein frame, matter doesn't fall along the geodesics of the metric, hence the energy-momentum conservation now reads
\be\label{en-mom-cons-Einstein}
\tilde{\nabla}_{\mu}(\tilde{T}^{\mu\nu}+t_{\phi}^{\mu\nu})=0
\ee
where $\tilde{T}^{\mu\nu}$ is the energy-momentum tensor for matter in Einstein frame and $t_{\phi}^{\mu\nu}$ is the one associated with the scalar field. Following the notation of section \ref{conformal}, we use a tilde to indicate quantities in the Einstein frame, (with the exception of the scalar field which doesn't have a counterpart in the Jordan frame).

In a similar fashion, Einstein equations include now terms dependent on the scalar field, as shown in eq.(\ref{einsteineom}). In particular, the terms on the RHS of eq.(\ref{einsteineom}) depending on the scalar field, correspond to the energy-momentum tensor of $\phi$, and hence we can rewrite eq.(\ref{einsteineom}) as follows
\be\label{Einstein-Einstein-frame}
\tilde{G}_{\mu\nu}=8\pi G(\tilde{T}_{\mu\nu}+t^{\phi}_{\mu\nu})
\ee
Similarly to what done in section IV, we can introduce small perturbations to the metric and the energy-momentum tensors, and derive the first order equations by expanding eq.(\ref{en-mom-cons-Einstein}) and eq.(\ref{Einstein-Einstein-frame}) in the perturbations \cite{Hwang-Noh}.  

Alternatively, we can obtain the same equations by mean of the mapping between Jordan and Einstein frame described in section III. In this section we will follow the latter method, and derive explicitily the mapping for the perturbed quantities from Jordan frame to the Einstein one. Once we have this mapping, it is straightforward to derive the perturbed equations by direct substitution into the equations derived in section IV.

In the Jordan frame, the perturbed metric is written as
\be
g_{\mu\nu}=g_{\mu\nu}^{(0)}+h_{\mu\nu} \ ,
\label{jordanpertmetric}
\ee
where the perturbation $h_{\mu\nu}$ is assumed small. Similarly, in the Einstein frame, the  perturbed metric
and scalar field are written as
\be
{\tilde g}_{\mu\nu}={\tilde g}_{\mu\nu}^{(0)}+{\tilde h}_{\mu\nu} \ \ \ \ \ \ \ \ \ \phi=\phi_0+\delta\phi Y\ ,
\label{einsteinpertmetric}
\ee
where $\delta\phi$ is a time dependent small perturbation and $Y=Y(\vec{x})$ is the set of harmonic functions. Using~(\ref{conftrans}), we can see that at first order
\be\label{perturbedmetric-relation}
\tilde{g}_{\mu\nu}=\tilde{g}^{(0)}_{\mu\nu}+\tilde{h}_{\mu\nu} = e^{\beta\kappa\phi_0}\left(g^{(0)}_{\mu\nu}+h_{\mu\nu}+\beta\kappa g_{\mu\nu}^{(0)}\delta\phi Y\right) \ .
\ee
Similarly, given the energy-momentum tensor $T^{\mu}_{\nu}$ in Jordan frame, we can use~(\ref{einsteinperfectfluid}) and~(\ref{conftrans}),  we can derive the following expression for the perturbation to the energy-momentum tensor of matter in Einstein frame
\be\label{perturbed-enmom-relation}
{\tilde{T}^{\mu}}_{\nu}={\tilde{T}^{\mu(0)}}_{\nu}+{\tilde{\delta{T}}^{\mu}}_{\nu}=e^{-2\beta\kappa\phi_0}\left({T^{\mu (0)}}_{\nu}+\delta {T^{\mu}}_{\nu}-2\beta\kappa {T^{\mu (0)}}_{\nu} \delta\phi Y\right).
\ee
Equations (\ref{perturbedmetric-relation}) and (\ref{perturbed-enmom-relation}), together with the background version of~(\ref{conftrans}) and~(\ref{einsteinperfectfluid}), are all is needed to map the perturbed expressions in Jordan frame into the ones in Einstein frame, as we show in the rest of this section.

\subsection{Metric perturbations}

Let's write the perturbed line element in Einstein frame as follows
\be\label{ds2_einstein}
d\tilde{s}^2=-\tilde{a}^2(\tau)(1+2\tilde{A}Y)d\tau^2-\tilde{a}^2(\tau)\tilde{B}Y_id\tau dx^i+\tilde{a}^2(\tau)(\gamma_{ij}+2\tilde{H}_LY\gamma_{ij}+2\tilde{H_T}Y_{ij})dx^idx^j
\ee
We do not need to consider vector and tensor perturbations as they cannot be generated through a conformal transformation.

Using~(\ref{perturbedmetric-relation}) and~(\ref{conftrans}) we can find the following explicit expressions for the metric elements
\ba
\tilde{g}_{00}&=&-\tilde{a}^2[1+2AY+\beta\kappa\delta\phi] \nonumber \\
\tilde{g}_{0j}&=&-\tilde{a}^2BY_j \nonumber \\
\tilde{g}_{ij}&=&\tilde{a}^2[\gamma_{ij}+(2H_L+\beta\kappa\delta\phi)\gamma_{ij}+2H_TY_{ij}] \nonumber \ .
\ea
Comparing the above expressions with (\ref{ds2_einstein}) we obtain the prescription to map the metric perturbations from Jordan to Einstein frame; specifically:
\ba\label{mapping-metric}
A&=&\tilde{A}-\frac{\beta}{2}\kappa\delta\phi \nonumber \\
B&=&\tilde{B}\nonumber\\
H_L&=&\tilde{H}_L-\frac{\beta}{2}\kappa\delta\phi \nonumber \\
H_T&=&\tilde{H}_T\ .
\ea

\subsection{Matter perturbations}

In the Einstein frame we can write  the first order perturbed energy-momentum tensor for matter, analogously to how we defined it in the Jordan frame (\ref{en-mom-tensor}), i.e.
\begin{eqnarray}\label{en-mom-einstein}
{{\tilde T}^0}_0&=&-\tilde{\rho}[1+ \tilde{\delta}\, Y],\nonumber\\
{\tilde{T}^0}_j&=&(\tilde{\rho}+\tilde{p})(\tilde{v}-\tilde{B})Y_j,\nonumber\\
{\tilde{T}^i}_j&=&\tilde{p}[\tilde{\delta}^i_j+\tilde{\pi}_LY\delta^i_j+\tilde{\pi}_TY^i_j].
\end{eqnarray}
Using~(\ref{perturbed-enmom-relation}) and~(\ref{einsteinperfectfluid}) we can find the following explicit expressions for its components
\ba
{{\tilde T}^0}_{0}&=&-\tilde{\rho}[1+\delta Y-2\beta\kappa\delta\phi Y] \nonumber \\
{{\tilde T}^0}_j&=&(\tilde{\rho}+\tilde{p})(v-B)Y_j \nonumber \\
{{\tilde T}^i}_j&=&\tilde{p}[{\delta^i}_j+(\pi_L-2\beta\kappa\delta\phi) Y{\delta^i}_j+\pi_TY^i_j]\nonumber \ .
\ea
Finally, comparing with (\ref{en-mom-einstein}) we get the following prescription to map the matter perturbations:
\ba\label{mapping-matter}
\delta&=&\tilde{\delta}+2\beta\kappa\delta\phi \nonumber \\
v&=&\tilde{v}\nonumber\\
\pi_L&=&\tilde{\pi}_L+2\beta\kappa\delta\phi \nonumber \\
\pi_T&=&\tilde{\pi}_T\ .
\ea
The set of mappings~(\ref{mapping-metric}) and~(\ref{mapping-matter}) is all we need to obtain the first order equations in Einstein frame directly from the equations in Jordan frame, which we derived explicitely in section IV.

\section{Einstein frame background evolution scaling behavior}
\label{scaling}
In this section, for completeness, we demonstrate the Einstein frame scaling behavior that occurs with the onset of non-negligible $f(R)$ coupling relevant in section \ref{evol}, as pointed out in \cite{Amendola:1999er}.  Consider the coupled equations (where the dot represents $d/d\tilde t$)
\bea
\dot{\tilde{\rho}}_{m}+3\tilde H\tilde\rho_{m} &=& -C\beta\kap \dot{\phi} \tilde\rho_{m} \la{meq}\\
\dot{\tilde\rho}_{\gam}+4\tilde H\tilde\rho_{\gam} &=& 0 \la{geq}\\
\ddot{\phi}+3\tilde H\dot{\phi}+V_{\phi} &=& C\beta\kap\tilde\rho_{m} \la{phieq}\\
H^{2} &=& \frac{\kap^{2}}{3}\le[\frac{\dot{\phi}^{2}}{2}+V+\tilde\rho_{m}+\tilde\rho_{\gam}\ri] \la{Heq}
\\
\dot{H} &=& -\frac{\kap^{2}}{2}\le[\dot{\phi}^{2}+\tilde\rho_{m}+\frac{4}{3}\tilde\rho_{\gam}\ri] \la{Hdeq}
\eea
taking $V\simeq A\exp(-\kap\beta\mu\phi)$ appropriate for many potentials $V(\phi)$ arising out of the conformal transformations of $f(R)$ theories for large values of $\phi$.
Following \cite{Amendola:1999er} we introduce the parameterization, 
\bea
x&=& \fr{\kap}{\tilde H}\fr{\dot{\phi}}{\sqrt{6}}, \ \ \ y = \fr{\kap}{\tilde H}\fr{\sqrt{V}}{\sqrt{3}}, \ \ \ z = \fr{\kap}{\tilde H}\fr{\sqrt{\tilde\rho_{\gam}}}{\sqrt{3}}, \ \ \ \alpha = \ln \tilde a
\eea
The Friedmann equation (\ref{Heq}) allows the matter density to be defined in terms of $x,y,z$
\bea
\frac{\kap^{2}}{\tilde H^{2}}\frac{\tilde\rho_{m}}{3} &=& 1-x^{2}-y^{2}-z^{2}
\eea
and 
\bea
\frac{\dot{\tilde H}}{\tilde H^{2}} 
&=& -\frac{3}{2}\le[1+x^{2}-y^{2}+\frac{z^{2}}{3}\ri]
\eea
Representing $d/d\alpha$ by ', equation (\ref{geq}) gives
\bea
z' &=& -z\le[1-3x^{2}+3y^{2}-z^{2}\ri]
\eea
In the presence of radiation
\bea
\fr{\dot{\tilde H}}{\tilde H^{2}}&=& -\le[\frac{z'}{z}+2\ri] \la{Hzeq}
\\
y' &=& -\mu yx+y\le[\frac{z'}{z}+2\ri]
\\
x'&=& x\le(\frac{z'}{z}-1\ri)+\mu y^{2} +C\le(1-x^{2}-y^{2}-z^{2}\ri)
\eea
If the radiation contribution is negligible one can't use (\ref{Hzeq}) and instead one gets
\bea
y' &=&  y\le[ -\mu x+\frac{3}{2}\le(1+x^{2}-y^{2}\ri)\ri] \la{yeq}\\
x' &=& C-\fr{3x}{2}-Cx^{2}+\frac{3x^{3}}{2}+y^{2}\le(\mu-C-\frac{3x}{2}\ri)\la{xeq}
\eea

Scaling attractors satisfy the constraint $x'=y'=z'=0$ thereby giving $\Omega_{\phi}$=const.
We are interested in the regime in which the $f(R)$ coupling (for which $C=1/2$) comes to be important in the matter dominated era. From (\ref{yeq}) and (\ref{xeq}) we see that the attractor requires $y=0$ and $x=1/3$ so that $\Omega_{\phi}=1/9$ and $w_{\phi}=1$ i.e. the scalar field is purely kinetic, and $w_{eff}=1/9$. In this case
\bea
\tilde{a}&\propto & \tilde{t}^{\frac{3}{5}}\propto\tau^{\frac{3}{2}} .
\eea
Considering the Friedmann equation, therefore, one finds a scalar field solution
 $\phi = \phi_{0}\ln \l(t/ t_{0}\r)$ with $\phi_{0}=3/\beta\kap$. Converting to the Jordan frame one obtains
\bea
t &=& \int \exp\l(-\frac{\beta\kap\phi}{2}\r) d\tilde{t} 
\propto  \tilde{t}^{\frac{4}{5}}\\
a &=& \exp\l(-\frac{\beta\kap\phi}{2}\r)\tilde{a}
\propto \tilde{t}^{\frac{2}{5}}
\propto t^{\frac{1}{2}}.
\eea

In the accelerating regime the Einstein frame evolution tends towards and attractor with $x=\lambda/\sqrt{6}$ and $y =\sqrt{ (1-\lambda^{2}/6)}$ and $\Omega_{c} \approx 0$ so that $w_{eff} = 1-\lambda^{2}/3$.

\end{document}